\documentclass[letterpaper,11pt]{article} 
\pdfoutput=1                                

\usepackage[T1]{fontenc}                    

\usepackage{jheppub}                        

\usepackage{mathtools}                      
\usepackage{amssymb,amsfonts}               
\usepackage{mathrsfs}                       
\usepackage{bm}                             

\usepackage{physics}                        

\usepackage{graphicx}                       
\usepackage{subcaption}                     

\usepackage{booktabs}                       
\usepackage{tabularx}                       
\usepackage{multirow}                       
\usepackage{makecell}                       
\usepackage[table]{xcolor}                  

\usepackage{tikz}                           
\usepackage{tikz-cd}                        
\usetikzlibrary{shapes.geometric,arrows,positioning} 

\usepackage{empheq}                         

\hypersetup{                                
  colorlinks=true,
  linkcolor=blue,
  citecolor=blue,
  urlcolor=blue
}

\newcommand*\widefbox[1]{\fbox{\hspace{2em}#1\hspace{2em}}}

\newcommand{\nn}{\nonumber}

\newcommand{\faddeeva}{\operatorname{w}}    

\title{Two-Photon Emission from an Accelerated Detector with Gaussian Switching}

\author[a]{Arash Azizi}
\affiliation[a]{{\it The Institute for Quantum Science and Engineering,
Texas A\&M University,\\ College Station, TX 77843, U.S.A.}}
\emailAdd{sazizi@tamu.edu}

\abstract{We present a closed-form analysis of second-order two-photon emission from a uniformly accelerated Unruh–DeWitt detector with a smooth Gaussian switch and derivative coupling. The time-ordered Dyson integrals are evaluated exactly, yielding a single spectral amplitude that simultaneously covers all four channels—right-travelling/right-travelling (RR), left-travelling/left-travelling (LL), right-travelling/left-travelling (RL), and left-travelling/right-travelling (LR). The spectra are organized by narrow kinematic gates (sum-frequency for RR/LL, difference-frequency for RL/LR) and a causal resonant denominator encoded by the Faddeeva function. As the gate time increases, the Gaussian factors sharpen into delta constraints and recover the eternal-interaction response. We map scaling with acceleration, gap, and duration, and show that the single-particle marginal combines a sharp nonthermal resonance with an Unruh thermal tail.
}

\begin{document}
\maketitle
\flushbottom

\section{Introduction}

Uniform acceleration imprints a thermal character on the Minkowski vacuum with Unruh temperature $T_U = \frac{a}{2\pi}$ tying together accelerated observers, black--hole thermodynamics, and cosmological particle production \cite{Parker1968,Fulling1973,Unruh1976,Davies1975,Hawking1974,Hawking1975}. 
A standard operational probe is the Unruh--DeWitt (UDW) detector, which models localized light--matter interactions in quantum field theory (QFT) and provides a clean bridge to quantum--optical intuition \cite{Unruh1976,Einstein100,Colosi2009Rovelli}. 
Much is known at the level of single--quantum excitation, switching regularization, and response rates \cite{Lin2006,Louko2006,Louko2008}.

Two--photon processes, however, probe phase--coherent energy exchange, causal time ordering, and vacuum entanglement in ways that single--photon observables cannot. 
From the perspective of line shapes and finite--time resonances, they naturally connect detector QFT with classic quantum field--theoretic frameworks due to Dirac, Weisskopf--Wigner, and Dyson \cite{Dirac1927,Wigner_Weisskopf1930,Dyson1949}. 
In relativistic quantum information (RQI), multi--quantum sectors underpin entanglement harvesting from the vacuum and related protocols \cite{Reznik2003,Pozas2015harvesting,Martin-Martinez2016harvesting,Simidzija2018}, and they interface with quantum--optical viewpoints that emphasize causality and entanglement structure in Unruh/Hawking settings \cite{Svidzinsky2021PRR,Svidzinsky2021PRL,Scully2022}.

This work develops an exact, unified description of the two--photon sector of a uniformly accelerated UDW detector with derivative coupling and a smooth, normalized Gaussian switch. 
We derive a closed--form two--photon spectral amplitude that simultaneously covers all four emission channels---right--travelling/right--travelling (RR), left--travelling/left--travelling (LL), right--travelling/left--travelling (RL), and left--travelling/right--travelling (LR).

The geometry of the spectrum is organized by narrow kinematic gates---sum--frequency gates for intra--wedge channels and difference--frequency gates for inter--wedge channels---and by a causal resonant denominator. 
The gates sharpen to $\delta$--function constraints as the interaction time grows, while the resonant structure is the same finite--time line shape that governs textbook decay problems, here captured compactly by the Faddeeva function. 
The formalism makes transparent how the single--particle marginal inherits a thermal Unruh tail together with a sharp, nonthermal resonance at the detector gap.

At the calculational core is the second--order Dyson series. 
The two--photon sector arises from a strictly time--ordered double integral $\int d\tau \int_{-\infty}^{\tau} d\tau'$ corresponding physically to a counter--rotating $\ket{g}\!\to\!\ket{e}$ excitation by vacuum fluctuations followed by a rotating--wave $\ket{e}\!\to\!\ket{g}$ decay that radiates the second quantum. 
In our derivative--coupling model, only the creation parts of the field evaluated on the accelerated worldline survive on the vacuum, which organizes the result cleanly into the four chirality channels. 
The Gaussian--gated Dyson integrals are exactly evaluable and produce the same causal denominator familiar from the Dirac--Weisskopf--Wigner--Dyson picture of finite--time emission and open--system line shapes \cite{Dirac1927,Wigner_Weisskopf1930,Dyson1949,Agarwal2012,BreuerPetruccione2002}, now transplanted into detector QFT \cite{UnruhWald1984,Lin2006,Louko2008}. 
The present treatment generalizes our recent closed--form analysis of uniformly accelerated two--photon emission to a single chirality--indexed amplitude that unifies all channels and exposes the distributional large--$T$ limit \cite{Azizi2025TPE}.

A smooth Gaussian switch is key for both physics and mathematics. 
Sudden switching seeds unphysical transients and regularization ambiguities in detector models, whereas a normalized Gaussian $s(\tau) = (T\sqrt{2\pi})^{-1} \exp\!\left[-\frac{\tau^2}{2T^2}\right]$ cleanly represents a finite--duration interaction, admits controlled long--time limits, and enables exact evaluation of the time--ordered integrals \cite{Louko2006,Louko2008,Lin2006}. 
In frequency space, the gate acts as a $\delta$--sequence that sharpens the kinematic constraints as $T\!\to\!\infty$, while the Faddeeva factor yields an analytic, causal line shape with algebraic tails. 
This choice also matches widespread practice in RQI, where Gaussian switching is routinely used to model localized probes and to control UV/IR behavior in harvesting and communication scenarios \cite{Reznik2003,Pozas2015harvesting,Martin-Martinez2016harvesting,Simidzija2018}.

Methodologically, our contributions are: 
(i) a normalized Gaussian--switch implementation that regulates time ordering at the amplitude level; 
(ii) a unified, chirality--indexed formula that treats RR/LL/RL/LR on equal footing; 
(iii) an explicit distributional map from finite--duration coupling to the eternal--interaction limit that reproduces the usual response while keeping the finite--time line shape visible; 
and (iv) a clear account of how the single--particle marginal merges a nonthermal resonance with an Unruh thermal tail. 
These results tie together detector theory and response regularization \cite{UnruhWald1984,Lin2006,Louko2008}, classic line--shape intuition \cite{Dirac1927,Wigner_Weisskopf1930,Dyson1949,Agarwal2012,BreuerPetruccione2002}, and the RQI perspective on vacuum correlations \cite{Reznik2003,Pozas2015harvesting,Martin-Martinez2016harvesting,Simidzija2018,Scully2022}.

The paper is organized as follows. 
Section~\ref{sec:framework} defines the model, the derivative coupling with Gaussian switching, and the Unruh--mode decomposition on the accelerated worldline. 
Section~\ref{sec:evolution} presents the second--order Dyson calculation and derives the unified two--photon spectral amplitude, analyzes its scaling, and extracts the distributional large--$T$ limit. 
Section~\ref{sec:viz} visualizes the full two--dimensional spectra, examines resonant cuts and single--particle marginals, and shows the emergence of Unruh thermality in the high--frequency tails. 
Technical details of the time--ordered integrals, distributional limits, and channel mapping are collected in the Appendices.

\section{Theoretical Framework} \label{sec:framework}
We consider a two-level Unruh-DeWitt (UDW) detector with ground state $|g\rangle$, excited state $|e\rangle$, and energy gap $\omega$. The detector moves with constant proper acceleration $a$ through a (1+1)-dimensional massless scalar field $\Phi$ in Minkowski spacetime. The primary goal of this work is to analyze the two-photon emission resulting from the detector's interaction with the field, starting from the initial state $|g\rangle \otimes |0_M\rangle$.

\subsection{Interaction Hamiltonian with Gaussian switching}

The dynamics of the system are governed by the interaction Hamiltonian in the interaction picture. To model a physically realistic interaction of finite duration, we introduce a smooth Gaussian switching function $s(\tau)$, where $\tau$ is the detector's proper time. The Hamiltonian is given by 
\begin{equation}
H_{\text{int}}(\tau)=g\,s(\tau)\,
\frac{\partial}{\partial\tau}\Phi(x(\tau))\,
\big(\sigma^\dagger e^{i\omega\tau}+\sigma e^{-i\omega\tau}\big),
\label{eq:hamiltonian}
\end{equation}
where $g$ is the coupling constant and $\sigma^\dagger, \sigma$ are the detector's raising and lowering operators. The derivative coupling ensures a well-behaved transformation to the accelerated frame. The Gaussian switch is normalized such that $\int s(\tau) d\tau = 1$:
\begin{equation}
s(\tau)=\frac{1}{T\sqrt{2\pi}}e^{-\tau^2/(2T^2)}.
\label{eq:switch}
\end{equation}
The parameter $T$ controls the duration of the interaction.

\subsection{Unruh mode expansion}

For a massless scalar field in $(1+1)$ dimensions, it is convenient to use the light-cone coordinates
\[
u = t - x, \qquad v = t + x,
\]
so that right-moving modes depend only on $u$ and left-moving modes only on $v$.  
Accordingly, we decompose the field into two independent chiral parts,
\[
\Phi(u,v) \;=\; \Phi_{\mathrm{RTW}}(u) + \Phi_{\mathrm{LTW}}(v).
\]
Unruh modes \cite{Unruh1976, UnruhWald1984} are specific combinations of Rindler modes from both spacetime wedges that are defined to have the Minkowski vacuum as their vacuum state.  They are
\begin{align}
\Phi_{\text{RTW}}(u)= \int_{-\infty}^{+\infty} d\Omega
\Bigg\{& \Big(
\theta(u)\, f(\Omega)\, u^{i\Omega}
+ \theta(-u)\, f(-\Omega)\, (-u)^{i\Omega} \Big) A_{\Omega} \nn\\
& + \Big(
\theta(u)\, f(\Omega)\, u^{-i\Omega}
+ \theta(-u)\, f(-\Omega)\, (-u)^{-i\Omega} \Big) A_{\Omega}^{\dagger}
\Bigg\}, 
\label{Unruh.mode.RTW}
\end{align}
and
\begin{align}
\Phi_{\text{LTW}}(v)= \int_{-\infty}^{+\infty} d\Omega
\Bigg\{& \Big(
\theta(v)\, f(\Omega)\, v^{i\Omega}
+ \theta(-v)\, f(-\Omega)\, (-v)^{i\Omega} \Big) B_{\Omega} \nn\\
& + \Big(
\theta(v)\, f(\Omega)\, v^{-i\Omega}
+ \theta(-v)\, f(-\Omega)\, (-v)^{-i\Omega} \Big) B_{\Omega}^{\dagger}
\Bigg\}, 
\label{Unruh.mode.LTW}
\end{align}
where the mode normalization is
\begin{equation}
f(\Omega) = \frac{e^{- \pi \Omega / 2}}{\sqrt{8\pi \Omega \sinh(\pi \Omega)}}.
\label{f}
\end{equation}
The associated Unruh operators satisfy canonical commutation relations,
\begin{equation}
[A_\Omega,A_{\Omega'}^\dagger]=\delta(\Omega-\Omega'),\qquad
[B_\Omega,B_{\Omega'}^\dagger]=\delta(\Omega-\Omega'),
\end{equation}
and annihilate the Minkowski vacuum,
\begin{equation}
A_\Omega\ket{0_M}=B_\Omega\ket{0_M}=0.
\end{equation}

\subsection{Unruh modes restricted to the accelerated worldline}

Along the uniformly accelerated worldline in wedge~I ($u<0,\;v>0$),
\[
u(\tau) = -\frac{1}{a}\,e^{-a\tau},\qquad
v(\tau) = \frac{1}{a}\,e^{a\tau},
\]
the Unruh-mode expansions restricted to the trajectory take the form
\begin{align}
\Phi_{\mathrm{RTW}}(\tau)
&= \int_{-\infty}^{+\infty}\! d\Omega\,
f(-\Omega)\Big(a^{-i\Omega} e^{-i a \Omega \tau}\, A_\Omega
\;+\; a^{i\Omega} e^{i a \Omega \tau}\, A_\Omega^\dagger\Big), \label{eq:phiRTW_tau}\\
\Phi_{\mathrm{LTW}}(\tau)
&= \int_{-\infty}^{+\infty}\! d\Omega\,
f(\Omega)\Big(a^{-i\Omega} e^{i a \Omega \tau}\, B_\Omega
\;+\; a^{i\Omega} e^{-i a \Omega \tau}\, B_\Omega^\dagger\Big), \label{eq:phiLTW_tau}
\end{align}
with $f(\Omega)$ the normalization in Eq.~\eqref{f}. The proper-time phases
$e^{\mp i a \Omega \tau}$ arise from evaluating the \emph{Minkowski}-positive-frequency
Unruh modes on the accelerated trajectory; they do not define positive frequency
with respect to $\tau$.

Since the field starts in the Minkowski vacuum, only the creation parts contribute when acting
on $\ket{0_M}$. We adopt the convention that the superscript $(-)$ denotes the creation part
(built from $A_\Omega^\dagger,B_\Omega^\dagger$). From Eqs.~\eqref{eq:phiRTW_tau}–\eqref{eq:phiLTW_tau},
\begin{align}
\Phi_{\mathrm{RTW}}^{(-)}(\tau)
&= \int_{-\infty}^{+\infty}\! d\Omega\,
f(-\Omega)\,a^{i\Omega} e^{i a \Omega \tau}\, A_\Omega^\dagger, \label{eq:phiRTW_create}\\
\Phi_{\mathrm{LTW}}^{(-)}(\tau)
&= \int_{-\infty}^{+\infty}\! d\Omega\,
f(\Omega)\,a^{i\Omega} e^{-i a \Omega \tau}\, B_\Omega^\dagger. \label{eq:phiLTW_create}
\end{align}
These are the only field pieces that survive on $\ket{0_M}$ in the Dyson expansion.
Terms with an annihilator to the left of a creator contract back to the vacuum and
do not populate the two–quantum sector; e.g.,
\[
A_\Omega A_{\Omega'}^\dagger \ket{0_M}
= \delta(\Omega-\Omega')\,\ket{0_M},
\]
which contributes only to the vacuum–to–vacuum (``bubble'') amplitude—an overall
phase/normalization that can be absorbed into the zeroth–order term. Since our
focus is on the emitted two–quantum state, we discard these vacuum–preserving
contributions in what follows.

\section{Time Evolution and the Two-Photon State} \label{sec:evolution}
We calculate the final two-photon state by beginning with the detector in its ground state $|g\rangle$ and the field in the Minkowski vacuum $|0_M\rangle$. The state evolution is governed by the second-order Dyson series, which captures the full $g \to e \to g$ emission process. This second-order contribution to the final state is
\begin{align}
\ket{\Psi_f} = \left( \frac{-i}{\hbar} \right)^2
\int_{-\infty}^{+\infty} d\tau \,s(\tau) g \frac{\partial \Phi}{\partial \tau} \sigma e^{-i \omega \tau}
\int_{-\infty}^{\tau} d\tau' \,s(\tau') g \frac{\partial \Phi}{\partial \tau'} \sigma^{\dagger} e^{i \omega \tau'}
\ket{0_M} \ket{g}.
\label{psi_f-tau}
\end{align}
In this expression, the detector is first excited at proper time $\tau'$ and later de-excites at time $\tau$. To evaluate this, we substitute the full field operator $\Phi = \Phi_{\text{RTW}} + \Phi_{\text{LTW}}$ at both vertices. \eqref{eq:phiRTW_tau} and \eqref{eq:phiLTW_tau} contribute. The product of the two field operators expands into four terms, corresponding to the RR, RL, LR, and LL emission channels. 

\begin{align}
\ket{\Psi_f} =& \left( \frac{-ig}{\hbar} \right)^2
\int_{-\infty}^{+\infty} d\tau \int_{-\infty}^{\tau} d\tau' s(\tau)s(\tau')
\int d\Omega \int d\Omega' \, e^{i\omega(\tau'-\tau)} a^{i(\Omega+\Omega')}   \label{psi_f.2} \\
&\times \Bigg[
\big(\partial_\tau\Phi_{\text{RTW}}^{(-)}(\tau)\big)_{\Omega} \big(\partial_{\tau'}\Phi_{\text{RTW}}^{(-)}(\tau')\big)_{\Omega'}
+ \big(\partial_\tau\Phi_{\text{RTW}}^{(-)}(\tau)\big)_{\Omega} \big(\partial_{\tau'}\Phi_{\text{LTW}}^{(-)}(\tau')\big)_{\Omega'} \nonumber \\
& \phantom{\times \Bigg[}
+ \big(\partial_\tau\Phi_{\text{LTW}}^{(-)}(\tau)\big)_{\Omega} \big(\partial_{\tau'}\Phi_{\text{RTW}}^{(-)}(\tau')\big)_{\Omega'}
+ \big(\partial_\tau\Phi_{\text{LTW}}^{(-)}(\tau)\big)_{\Omega} \big(\partial_{\tau'}\Phi_{\text{LTW}}^{(-)}(\tau')\big)_{\Omega'}
\Bigg] \ket{0_M}\ket{g}.
\nonumber
\end{align}
Each of the four terms in the square brackets contains a time-ordered double integral over $\tau$ and $\tau'$, which can be evaluated exactly. The general form of these integrals is
\begin{equation}
\mathcal{I}(\alpha, \beta) = \int_{-\infty}^{\infty}\!d\tau\!\int_{-\infty}^{\tau}\!d\tau'\, s(\tau)s(\tau')\, e^{i\alpha\tau}\, e^{i\beta\tau'},
\end{equation}
which, upon a change of variables to $t=(\tau+\tau')/2$ and $s=\tau-\tau'$, evaluates to
\begin{equation}
\mathcal{I}(\alpha, \beta) = \frac{1}{2} \exp\left[-\frac{T^2}{4}(\alpha+\beta)^2\right] \faddeeva\left(\frac{T}{2}(\alpha-\beta)\right),
\end{equation}
where 
\begin{align}
    \faddeeva(z) = e^{-z^2}\text{erfc}(-iz),
\end{align}
is the Faddeeva function. By identifying the phases $\alpha$ and $\beta$ for each channel, we can determine their respective spectral amplitudes, $\mathcal{M}_{\chi\chi'}(\Omega, \Omega')$. For example, for the RR channel, $\alpha = a\Omega-\omega$ and $\beta = a\Omega'+\omega$. Combining the integral results with the prefactors from the field derivatives and normalizations yields the specific amplitudes.

The complete final state is the coherent sum of these four processes. After carrying out the integration and algebraic simplification for each channel, we can write the total final state in a compact form:
\begin{align}
\ket{\Psi_f} = \int_{-\infty}^{+\infty} d\Omega \int_{-\infty}^{+\infty} d\Omega' \Big(
&\mathcal{M}_{RR}(\Omega, \Omega') A^\dagger_\Omega A^\dagger_{\Omega'}
+ \mathcal{M}_{LL}(\Omega, \Omega') B^\dagger_\Omega B^\dagger_{\Omega'} \nonumber \\
 &+ \mathcal{M}_{RL}(\Omega, \Omega') A^\dagger_\Omega B^\dagger_{\Omega'}
+ \mathcal{M}_{LR}(\Omega, \Omega') B^\dagger_\Omega A^\dagger_{\Omega'}
\Big) \ket{0_M}\ket{g},
\label{psi_f.final}
\end{align}
where the exact, finite-time spectral amplitudes are given by:
\begin{align}
\mathcal{M}_{RR}(\Omega, \Omega') &= \frac{g^2 a^2}{2\hbar^2} \Omega\Omega' a^{i(\Omega+\Omega')} f(-\Omega)f(-\Omega') e^{-\frac{a^2T^2}{4}(\Omega+\Omega')^2} \faddeeva\left(T\left[ \frac{a}{2}(\Omega-\Omega')-\omega \right]\right), \nn\\
\mathcal{M}_{LL}(\Omega, \Omega') &= \frac{g^2 a^2}{2\hbar^2} \Omega\Omega' a^{i(\Omega+\Omega')} f(\Omega)f(\Omega') e^{-\frac{a^2T^2}{4}(\Omega+\Omega')^2} \faddeeva\left(T\left[-\frac{a}{2}(\Omega-\Omega')-\omega \right]\right), \nn\\
\mathcal{M}_{RL}(\Omega, \Omega') &= -\frac{g^2 a^2}{2\hbar^2} \Omega\Omega' a^{i(\Omega+\Omega')} f(-\Omega)f(\Omega') e^{-\frac{a^2T^2}{4}(\Omega-\Omega')^2} \faddeeva\left(T\left[ \frac{a}{2}(\Omega+\Omega')-\omega \right]\right), \nn\\
\mathcal{M}_{LR}(\Omega, \Omega') &= -\frac{g^2 a^2}{2\hbar^2} \Omega\Omega' a^{i(\Omega+\Omega')} f(\Omega)f(-\Omega') e^{-\frac{a^2T^2}{4}(\Omega-\Omega')^2} \faddeeva\left(T\left[-\frac{a}{2}(\Omega+\Omega')-\omega \right]\right). 
\end{align}
Equation \eqref{psi_f.final} is the central result of this paper, providing the complete two-photon state for a smoothly switched interaction.

\subsection{Lineshape and size of the signal}

From the exact identity (Appendix~\ref{app:limits-Gauss}, Eq.~\eqref{eq:I-exact}), 
the finite-time causal lineshape is carried by the Faddeeva factor
\begin{equation}
\faddeeva\!\Big(T\,\Delta_{\chi\chi'}\Big),\qquad 
\Delta_{\chi\chi'}\equiv \frac{a}{2}\big(\chi\Omega-\chi'\Omega'\big)-\omega.
\end{equation}
It is $\mathcal{O}(1)$ only near $\Delta_{\chi\chi'}\approx 0$ and has algebraic tails
\begin{equation}
\big|\faddeeva(T\Delta)\big|\sim \frac{1}{\sqrt{\pi}\,T\,|\Delta|},\qquad |T\Delta|\gg 1,
\end{equation}
so the resonance width is $\sim T^{-1}$. In the short-gate limit, the small-$T$ expansion 
(Appendix, Eq.~\eqref{eq:I-smallT}) gives
\begin{equation}
\mathcal{I}(\alpha,\beta)
=\frac{1}{2}+\frac{iT}{2\sqrt{\pi}}(\alpha-\beta)
-\frac{T^2}{8}\big[(\alpha-\beta)^2+(\alpha+\beta)^2\big]+\mathcal{O}(T^3),
\end{equation}
which tends to the instantaneous, time-ordered value $\tfrac12$ with controlled corrections.

Crucially, the Unruh weights do not always force exponential suppression. Using
\begin{equation}
f(-\chi\Omega)\,f(-\chi'\Omega')
=\frac{\exp\!\big[\tfrac{\pi}{2}(\chi \Omega+ \chi' \Omega')\big]}
{8\pi\,\sqrt{|\Omega\,\Omega'|\;\sinh(\pi|\Omega|)\,\sinh(\pi|\Omega'|)}},
\end{equation}
one sees that on the long-time kinematic gates (RR/LL: $\Omega'=-\Omega$; RL/LR: $\Omega'=\Omega$) 
the numerator’s exponential cancels, giving
\begin{equation}
f(-\Omega)\,f(+\Omega)=\frac{1}{8\pi\,|\Omega|\,\sinh(\pi|\Omega|)}.
\end{equation}
As $|\Omega|\to 0$, $\sinh(\pi|\Omega|)\sim \pi|\Omega|$ and 
$f(-\Omega)f(+\Omega)\sim (8\pi^2|\Omega|^2)^{-1}$ looks large; however, with derivative coupling 
the Dyson kernel contributes $(i\chi a\Omega)(i\chi' a\Omega')\propto \Omega\,\Omega'$, which on 
the gates gives $\Omega\,\Omega'=\mp \Omega^2$ and cancels the $1/|\Omega|^2$. The small-$|\Omega|$ 
behavior is therefore IR-safe, leaving a channel-independent constant prefactor.

At the resonant boost frequency $|\Omega|\simeq \omega/a$, two regimes appear:
\begin{enumerate}
\item Large gap / small acceleration ($\omega/a\gg1$): 
$\sinh(\pi|\Omega|)\approx \tfrac12 e^{\pi|\Omega|}$, so the amplitude acquires 
$\sim e^{-\pi\omega/a}$ and the probability $\sim e^{-2\pi\omega/a}$ (Unruh suppression).
\item Small gap / large acceleration ($\omega/a\lesssim1$): no exponential penalty; 
the overall size is set by the finite-time lineshape (through $T$), the IR-finite constant above, 
and the coupling $g$.
\end{enumerate}
In summary, $\faddeeva$ provides the narrow, causal line (width $\sim T^{-1}$, algebraic tails), and Unruh suppression only appears when $|\Omega|\sim \omega/a$ is large. When 
$f(-\chi\Omega)f(-\chi'\Omega')$ is $\mathcal{O}(1)$ (or large as $|\Omega|\to0$), 
the derivative coupling renders the amplitude finite; its magnitude is then controlled by 
$g$ and the switching window rather than by a thermal Boltzmann factor.

\subsection{\texorpdfstring{Asymptotic behavior as $T\to\infty$}{}}

The large-$T$ limits must be taken in the distributional sense 
(Appendix~\ref{app:limits-Gauss}). Writing $\sigma=\alpha+\beta$ and 
$\Delta=\alpha-\beta$, we have (Appendix, Eqs.~\eqref{eq:deltaT}--\eqref{eq:I-largeT-asymp})
\begin{align}
\delta_T(\sigma)&=\frac{T}{2\sqrt{\pi}}e^{-\tfrac{T^2}{4}\sigma^2}
\xrightarrow[T\to\infty]{\ \mathcal S'\ }\delta(\sigma), \\
\frac{1}{2}\,\faddeeva\!\Big(\tfrac{T}{2}\Delta\Big)&=
\frac{i}{\sqrt{\pi}\,T}\,\frac{1}{\Delta+i0^{+}}+O(T^{-3}), \\
\mathcal{I}(\alpha,\beta)&=
\frac{2i}{T^{2}}\,\delta_T(\sigma)\,\frac{1}{\Delta+i0^{+}}+O(T^{-4}).
\end{align}
Thus the strict distributional limit is zero:
\begin{equation}
\mathcal{I}(\alpha,\beta)\xrightarrow[T\to\infty]{\ \mathcal S'\ }0.
\end{equation}
However, if one extracts the natural $\delta$–sequence and considers the rate-type kernel
\begin{align}
P_T(\sigma,\Delta)
&\equiv \delta_T(\sigma)\,\frac{1}{2}\,\faddeeva\!\Big(\tfrac{T}{2}\Delta\Big)
\xrightarrow[T\to\infty]{\ \mathcal S'\ } 0,\\
T\,P_T(\sigma,\Delta)
&\xrightarrow[T\to\infty]{\ \mathcal S'\ }
\delta(\sigma)\,\frac{i}{\sqrt{\pi}}\,\frac{1}{\Delta+i0^{+}}.
\end{align}
This is the familiar infinite-time causal structure. In our variables,
\begin{equation}
\sigma=a(\chi\Omega+\chi'\Omega'),\qquad
\Delta=a(\chi\Omega-\chi'\Omega')-2\omega,
\end{equation}
so $\delta(\sigma)=\delta\!\big(a(\chi\Omega+\chi'\Omega')\big)=\frac{1}{|a|}\,\delta\!\big(\chi\Omega+\chi'\Omega'\big)$
(if $a>0$, replace $1/|a|$ by $1/a$).

So the Gaussian gate enforces the RR/LL sum-frequency constraint 
$\Omega'=-\Omega$ and the RL/LR difference-frequency constraint 
$\Omega'=\Omega$, while the Faddeeva factor produces the causal 
denominator in the detuning variable. Substituting this rate-type 
kernel into the general amplitude and integrating over $\Omega'$ 
reproduces the standard eternal-interaction result, whereas the 
unrescaled amplitude itself vanishes in the strict $T\to\infty$ limit.

\section{Visualization of Two–Photon Spectra}
\label{sec:viz}
We now visualize the exact, finite-time spectral amplitudes to understand the rich physics encoded in the two-photon final state. We begin by examining the two-particle correlations that define the overall spectral geometry and its dependence on physical parameters. We then transition to a quantitative analysis of the resonant lineshape and conclude by studying the single-particle spectrum, which reveals the emergence of Unruh thermality.
\begin{figure}[h!]
    \centering
    \includegraphics[width=\textwidth]{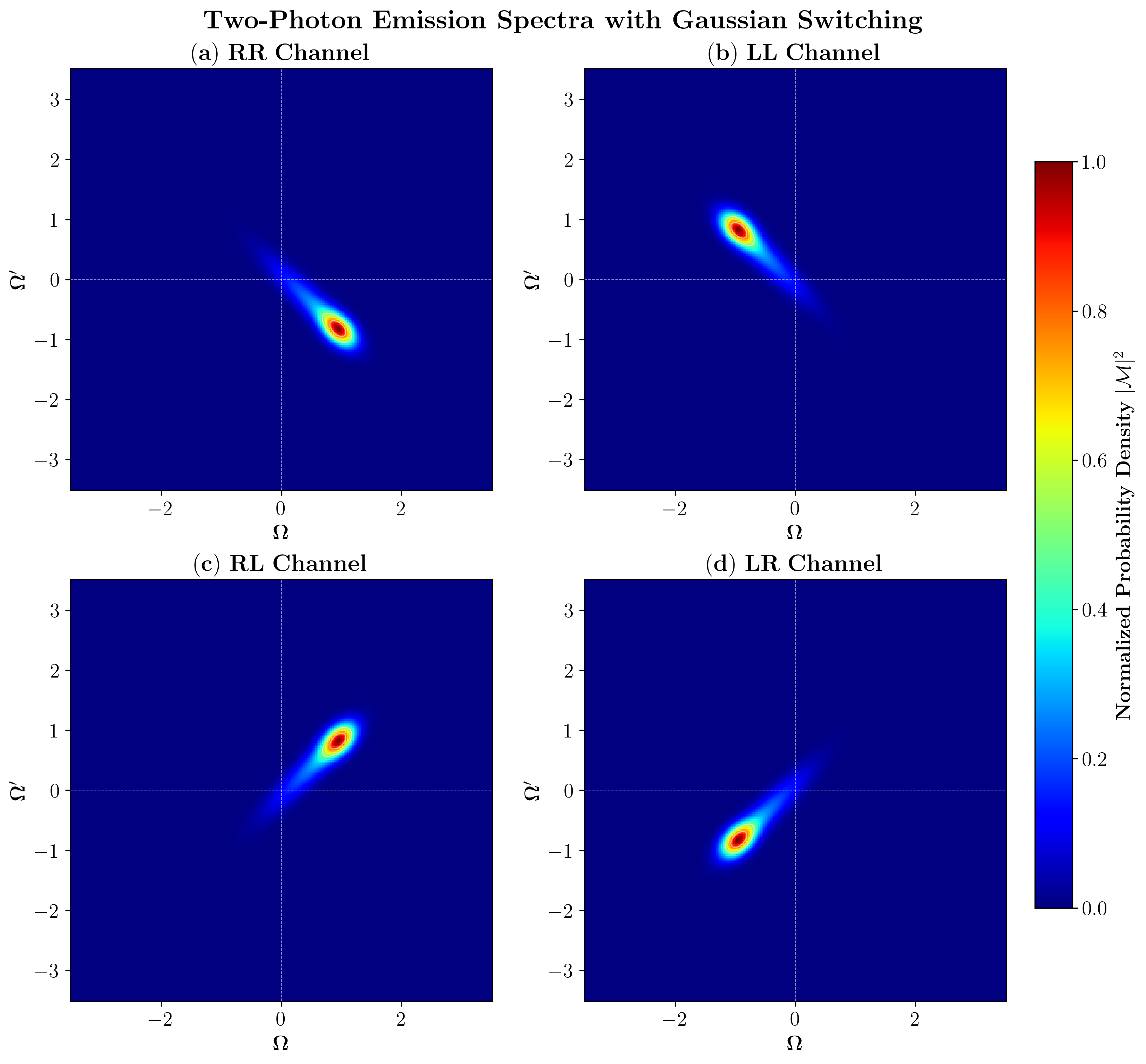}
    \caption{\textbf{Channel-resolved spectra at fixed $a=1$, $T\omega=5$.} Each panel shows the panel-normalized density $|\mathcal{M}_{\chi\chi'}(\Omega,\Omega')|^2$ (each panel independently normalized). Intra-wedge channels (RR, LL) concentrate along $\Omega'=-\Omega$, while inter-wedge channels (RL, LR) concentrate along $\Omega'=\Omega$. The bright hotspots correspond to the detector's resonant energy exchange.}
    \label{fig:channels}
\end{figure}
\subsection{Two-Particle Correlations and Spectral Geometry}
The most direct visualization of the emission process is the two-dimensional probability density $|\mathcal{M}_{\chi\chi'}(\Omega,\Omega')|^2$. Figure~\ref{fig:channels} displays these spectra for all four emission channels at a fixed acceleration and interaction time. Two primary features immediately stand out:
\begin{enumerate}
    \item Kinematic Gates: The emission is highly structured and constrained to specific kinematic lines. The intra-wedge channels (RR and LL) show a strong correlation along the sum-frequency line $\Omega' = -\Omega$, while the inter-wedge channels (RL and LR) are correlated along the difference-frequency line $\Omega' = \Omega$. These gates arise from the Gaussian factor in the amplitude and tend to $\delta$-function constraints as $T \to \infty$, enforcing the eternal-limit energy relations.
    \item Resonant Peaks: Along these gates, the emission is brightest where the Faddeeva argument is near zero, corresponding to the detector's resonant energy exchange. For the RR and RL channels, this peak occurs near $\Omega \approx \omega/a$, while for the LL and LR channels it occurs near $\Omega \approx -\,\omega/a$.
\end{enumerate}
\begin{figure}[h!]
    \centering
    \includegraphics[width=\textwidth]{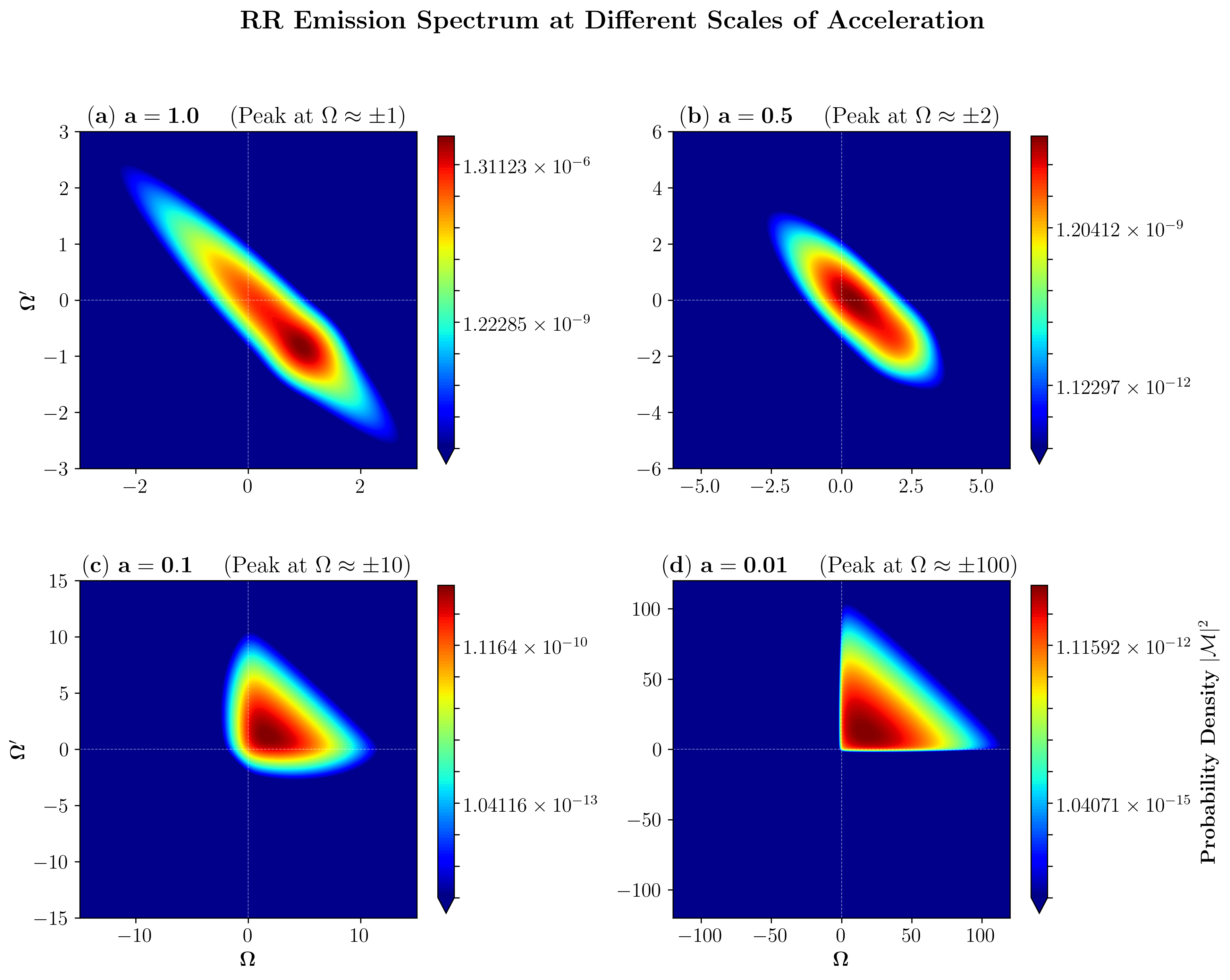}
    \caption{\textbf{Scaling with acceleration at fixed interaction time.} The RR channel spectrum is shown for decreasing acceleration $a$ at fixed $T\omega=5$. As $a$ decreases, the resonant peak follows $\Omega \simeq \omega/a$, shifting to higher frequencies. The colormap is logarithmic with the range set per panel for clarity.}
    \label{fig:multiscale-a}
\end{figure}

\begin{figure}[h!]
    \centering
    \includegraphics[width=\textwidth]{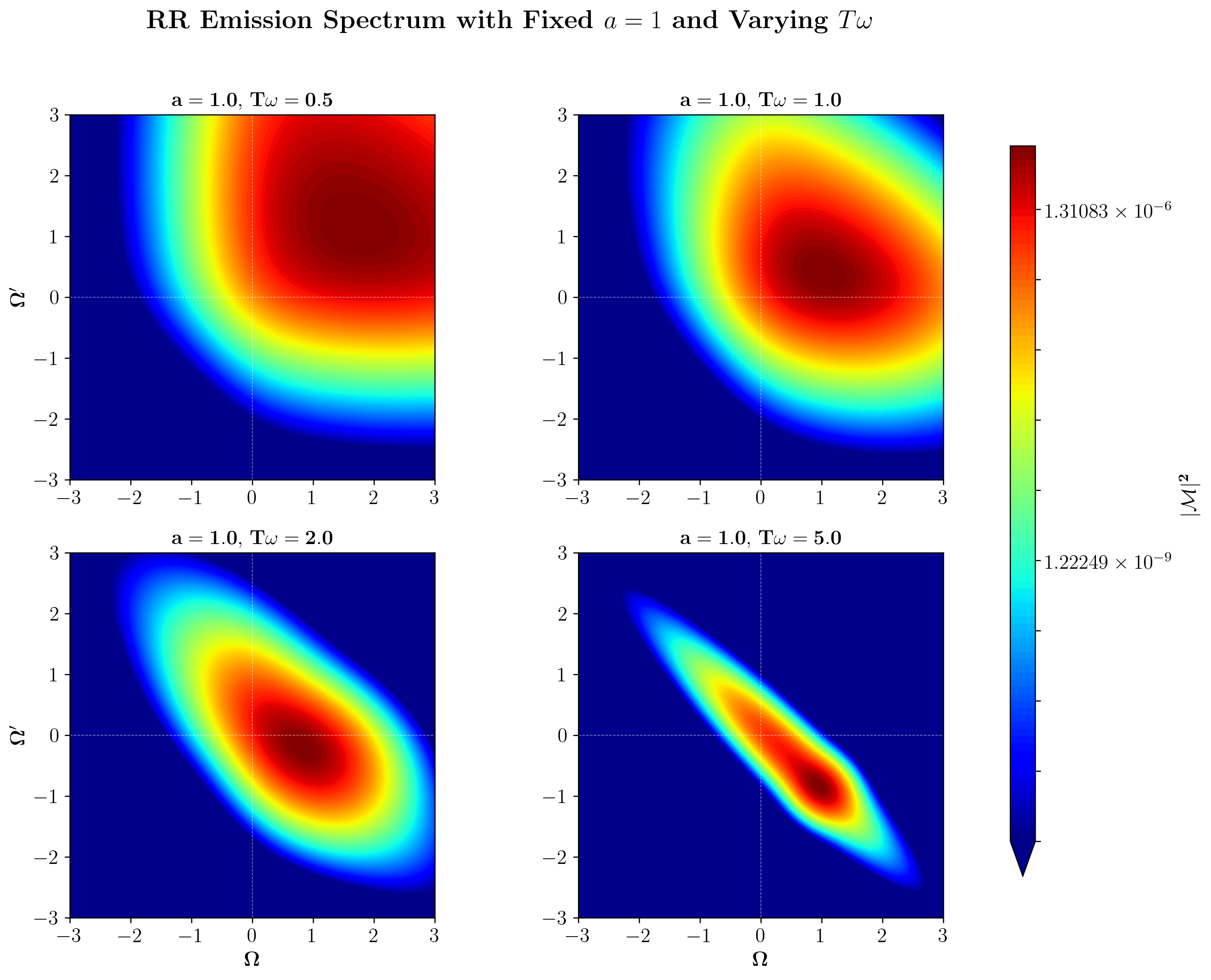}
    \caption{\textbf{Scaling with interaction time at fixed acceleration.} The RR channel spectrum is shown for increasing interaction time $T\omega$ at fixed $a=1$. As $T\omega$ increases, the sum-frequency gate along $\Omega'=-\Omega$ becomes progressively sharper, approaching the $\delta$-gate of the eternal limit.}
    \label{fig:RR-Tomega}
\end{figure}
These spectral features scale predictably with acceleration $a$ and interaction time $T$. As shown in Fig.~\ref{fig:multiscale-a} and Fig.~\ref{fig:RR-Tomega}, decreasing the acceleration pushes the resonant peaks to higher frequencies ($\Omega \sim \omega/a$), while increasing the interaction time sharpens the kinematic gates ($\text{width} \sim 1/T$), approaching the $\delta$-gated constraints of the eternal limit.

\subsection{Resonant Lineshape and Single-Particle Spectrum}
\begin{figure}[h!]
    \centering
    \begin{subfigure}[t]{0.49\textwidth}
        \centering
        \includegraphics[width=\textwidth]{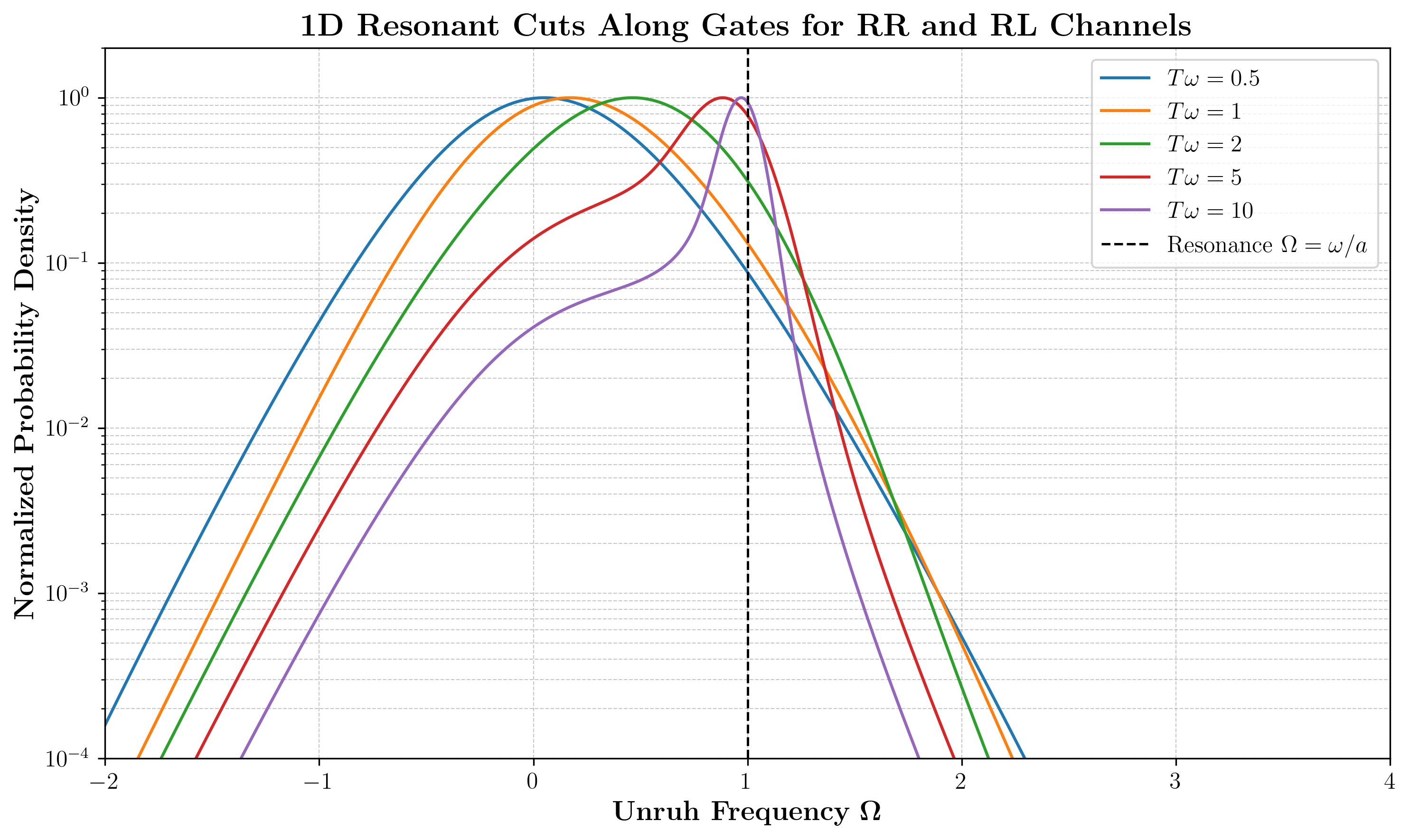}
        \caption{1D resonant cuts along the kinematic gates.}
        \label{fig:1d_resonant_cuts}
    \end{subfigure}
    \hfill
    \begin{subfigure}[t]{0.49\textwidth}
        \centering
        \includegraphics[width=\textwidth]{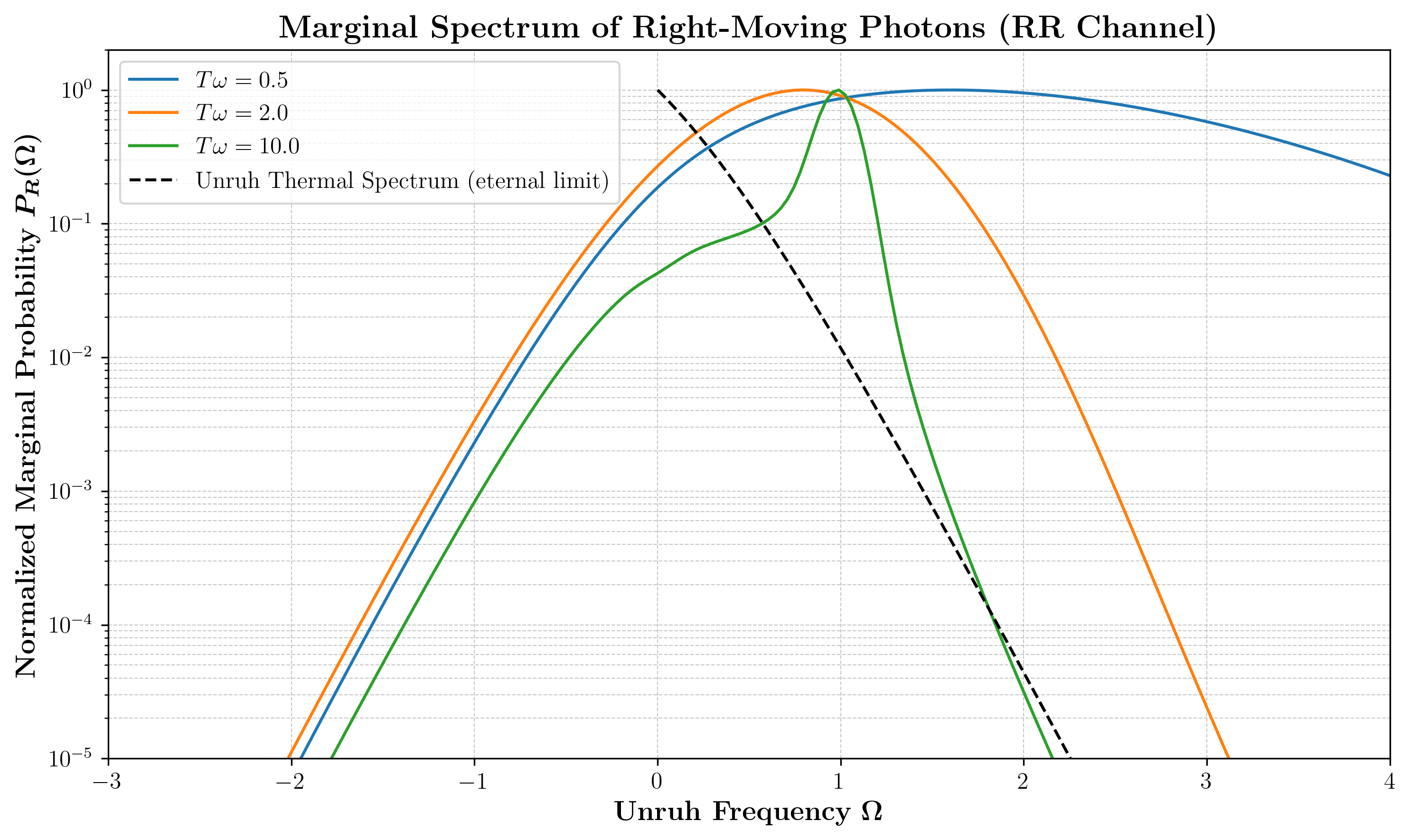}
        \caption{Marginal spectrum for a single right-moving photon.}
        \label{fig:marginal_spectra}
    \end{subfigure}
    \caption{\textbf{Quantitative analysis of the emission spectrum.} (a) Cuts along the resonant gates for RR/RL channels show the lineshape sharpening with increasing $T\omega$. (b) The marginal spectrum reveals the emergence of the Unruh thermal tail (dashed line) for long interaction times, alongside the detector's resonant peak at $\Omega=\omega/a$.}
    \label{fig:1d_plots}
\end{figure}

To analyze the structure of the emission more quantitatively, we examine one-dimensional slices and projections of the 2D spectra. First, by taking a cut directly along the kinematic gate (Fig.~\ref{fig:1d_resonant_cuts}), we resolve the resonant lineshape. As $T\omega$ increases, the peak sharpens and approaches the eternal-limit resonance at $\Omega=\omega/a$, with a width scaling as $\Delta\Omega\sim 1/T$.
Second, integrating over one photon's frequency yields the single-particle marginal (Fig.~\ref{fig:marginal_spectra}). For long interaction times ($T\omega=10$), the spectrum is hybrid: it shows a sharp, non-thermal peak at $\Omega=\omega/a$ together with a high-frequency tail that converges asymptotically to the Unruh thermal form $\propto \big(e^{2\pi\Omega}-1\big)^{-1}$ corresponding to $T_U=a/(2\pi)$. At short times, finite-time broadening generates noticeable off-resonant support. For clarity, the physical Rindler frequency domain is $\Omega\ge0$; any apparent weight at $\Omega<0$ in extended plots reflects the finite-time kernel’s even extension rather than negative-frequency quanta.
For reference, the key parameters and limiting behaviors for all four channels are summarized in Table~\ref{tab:channel-map}.
\begin{table}[h!]
\centering
\rowcolors{2}{gray!20}{white}
\renewcommand{\arraystretch}{1.8}
\setlength{\tabcolsep}{5pt}
\begin{tabular}{@{} l c l c l l @{}}
\toprule
\rowcolor{gray!40}
\textbf{Channel} & \textbf{\makecell{Chirality \\ $(\chi,\chi')$}} & \textbf{Integrand args. $(\alpha,\beta)$} & \textbf{\makecell{Gaussian gate \\ ($T\!\to\!\infty$)}} & \textbf{\makecell{$w$-argument \\ ($T\!\to\!\infty$)}} & \textbf{\makecell{Denominator \\ (from $\Delta+i0^+$)}} \\
\midrule
RR & $(+1,+1)$ & $\big(a\Omega-\omega,\; a\Omega'+\omega\big)$ & $\Omega'=-\Omega$ & $T(a\Omega-\omega)$ & $\displaystyle \frac{1}{\Omega-\frac{\omega}{a} + i0^+}$ \\[6pt]
LL & $(-1,-1)$ & $\big(-a\Omega-\omega,\; -a\Omega'+\omega\big)$ & $\Omega'=-\Omega$ & $T(-a\Omega-\omega)$ & $\displaystyle \frac{1}{-\Omega-\frac{\omega}{a} + i0^+}$ \\[6pt]
RL & $(+1,-1)$ & $\big(a\Omega-\omega,\; -a\Omega'+\omega\big)$ & $\Omega'=\Omega$ & $T(a\Omega-\omega)$ & $\displaystyle \frac{1}{\Omega-\frac{\omega}{a} + i0^+}$ \\[6pt]
LR & $(-1,+1)$ & $\big(-a\Omega-\omega,\; a\Omega'+\omega\big)$ & $\Omega'=\Omega$ & $T(-a\Omega-\omega)$ & $\displaystyle \frac{1}{-\Omega-\frac{\omega}{a} + i0^+}$ \\
\bottomrule
\end{tabular}
\caption{Channel map for a ground-state initial detector $|g\rangle$. In the constant-coupling limit, RR/LL enforce the sum-frequency gate $\Omega'=-\Omega$, whereas RL/LR enforce the difference-frequency gate $\Omega'=\Omega$. The “Denominator” column is the large-$T$ causal factor obtained from $1/(\Delta+i0^+)$ after imposing the gate.}
\label{tab:channel-map}
\end{table}

\section{Conclusions and Outlook} \label{sec:conclusion}

We have presented an exact, finite-time analysis of two-photon emission from a uniformly accelerated Unruh–DeWitt detector with derivative coupling and a smooth, normalized Gaussian switch. Working to second order in the Dyson series, we derived a closed-form two–photon spectral amplitude that simultaneously covers all four emission channels—right-travelling/right-travelling (RR), left-travelling/left-travelling (LL), right-travelling/left-travelling (RL), and left-travelling/right-travelling (LR)—within a single chirality-indexed formula. The time-ordered double integral at the heart of the calculation is evaluated exactly and expressed through the Faddeeva function \(\faddeeva\), yielding a compact resonant kernel that makes causality, finite-time lineshapes, and their limiting behavior explicit.

A key structural outcome is the emergence of narrow kinematic gates: sum-frequency gates for intra-wedge channels and difference-frequency gates for inter-wedge channels. These gates sharpen as the interaction duration \(T\) increases and tend to \(\delta\)-function constraints in the distributional large-\(T\) limit. Along the gates, the Faddeeva factor supplies the causal denominator familiar from Dirac–Weisskopf–Wigner–Dyson line-shape theory, thereby providing a transparent bridge between detector quantum field theory and textbook resonance physics. The derivative coupling ensures infrared safety by canceling apparent small-\(|\Omega|\) divergences in the Unruh weights, so the overall signal scale is controlled by the finite-time lineshape and the coupling rather than by spurious switching artefacts.

The formalism offers a clean map between finite-duration coupling and the idealized eternal-interaction limit. Although the unrescaled amplitude vanishes strictly as \(T\to\infty\) in the sense of distributions, the associated rate-type kernel obtained by extracting the natural \(\delta\)-sequence reproduces the standard causal structure and energy constraints of an eternally accelerated detector. In this way, finite-time physics—gating, resonance width \(\sim T^{-1}\), and algebraic tails—is kept manifest while recovering the established eternal response when appropriate.

On observables, the two-dimensional spectra reveal bright resonant hotspots pinned to the gates, whose positions track the boost frequency \(\Omega\simeq \omega/a\). The single-particle marginal acquires a hybrid character: a sharp nonthermal peak near \(\omega/a\) coexisting with a high-frequency Unruh tail set by the KMS condition. This hybrid structure is a concise diagnostic of how coherent, detector-specific resonance and thermal, horizon-induced fluctuations coexist in a single measurement record.

Several natural extensions follow from the present framework. One may incorporate detector backreaction and higher orders in the coupling to quantify linewidth renormalization and possible multi-quantum coherences; include boundaries or cavities to study Purcell-like shaping of the kinematic gates; generalize to \(3{+}1\) dimensions, massive fields, or electromagnetic couplings; and analyze mixed-state fields or accelerated detectors in curved spacetimes to probe the robustness of the gate–resonance picture. From a phenomenological angle, the exact finite-time amplitude provides concrete targets for analogue platforms—moving-mirror analogues, cold-atom horizons, and circuit QED—where controllable switching and strong light–matter coupling can enhance acceleration signatures.

In summary, the combination of a smooth Gaussian switch with an exact second-order Dyson treatment yields a unified, closed-form description of two-photon emission by accelerated detectors. It captures, in a single analytic expression, the interplay of kinematic gating, causal resonance, and thermal detailed balance, and it furnishes a versatile baseline for both foundational studies and experimental explorations of acceleration-induced quantum radiation.

\appendix 

\section{Unified Two-Photon Amplitude with Gaussian Switching} \label{app:chi-formalism}
Here, we derive the unified two-photon spectral amplitude, $\mathcal{M}_{\chi\chi'}(\Omega,\Omega')$, starting from the second-order Dyson series. The final state of the field is given by
\begin{equation}
|\Psi_{f}\rangle=\left(\frac{-i}{\hbar}\right)^{2}\!\int_{-\infty}^{+\infty}\!d\tau \int_{-\infty}^{\tau}\!d\tau^{\prime}\,
H_{\text{int}}(\tau)\, H_{\text{int}}(\tau')\, |0_{M}\rangle|g\rangle.
\end{equation}
The two-photon emission requires the detector to be excited at the earlier time $\tau'$ and de-excite at the later time $\tau$. This selects the $\sigma^\dagger$ term from $H_{\text{int}}(\tau')$ and the $\sigma$ term from $H_{\text{int}}(\tau)$, fixing the detector's evolution as $|g\rangle \to |e\rangle \to |g\rangle$. This process also extracts the temporal phase factor $e^{i\omega(\tau'-\tau)}$ from the detector operators.

Since the field begins in the vacuum, only the creation parts of the field operators, $\Phi^{(-)}$, contribute. To treat all four emission channels (RR, RL, LR, LL) simultaneously, we introduce a chirality index $\chi\in\{+1,-1\}$ (for RTW/LTW) and define the unified creation operators $C_{+1,\Omega}^\dagger \equiv A_\Omega^\dagger$ and $C_{-1,\Omega}^\dagger \equiv B_\Omega^\dagger$. The corresponding unified worldline operator is
\begin{equation}
\partial_\tau\Phi_{\chi}^{(-)}(\tau)
= \int_{-\infty}^{\infty}\! d\Omega\,\mathcal{N}(\Omega)\,e^{\frac{\chi\pi\Omega}{2}}\,
\big(i\chi a\Omega\big)\,a^{i\Omega}\,e^{i\chi a\Omega\tau}\,C_{\chi,\Omega}^\dagger,
\end{equation}
with $\mathcal{N}(\Omega)=1/\sqrt{8\pi\,\Omega\,\sinh(\pi\Omega)}$.

Substituting two of these operators into the Dyson series for a specific channel $(\chi, \chi')$, the final state takes the form
\begin{equation}
|\Psi_{\chi\chi'}\rangle
= \int_{-\infty}^{\infty}\! d\Omega \int_{-\infty}^{\infty}\! d\Omega'\;
\mathcal{M}_{\chi\chi'}(\Omega,\Omega')\;
C_{\chi,\Omega}^\dagger\,C_{\chi',\Omega'}^\dagger\,|0_M\rangle\,|g\rangle.
\end{equation}
The spectral amplitude $\mathcal{M}_{\chi\chi'}$ is found by collecting all non-temporal factors. This separates the amplitude into mode-dependent prefactors and a time-ordered integral:
\begin{align}
\mathcal{M}_{\chi\chi'}(\Omega,\Omega')
&= \left(\frac{-ig}{\hbar}\right)^{2}
\big(i\chi a\Omega\big)\big(i\chi' a\Omega'\big)\,
a^{i(\Omega+\Omega')}\,\mathcal{N}(\Omega)\mathcal{N}(\Omega')\,
e^{\frac{\chi\pi\Omega+\chi'\pi\Omega'}{2}}\;
\mathcal{I}^{\chi\chi'}_{T}(\Omega,\Omega') \nonumber \\
&= \frac{g^2 a^2}{\hbar^2}\, \chi\chi'\Omega\Omega'\,
a^{i(\Omega+\Omega')}\,\mathcal{N}(\Omega)\mathcal{N}(\Omega')\,
e^{\frac{\chi\pi\Omega+\chi'\pi\Omega'}{2}}\;
\mathcal{I}^{\chi\chi'}_{T}(\Omega,\Omega'),
\end{align}
where the integral over the switching functions and temporal phases is
\begin{equation}
\mathcal{I}^{\chi\chi'}_{T}(\Omega,\Omega')
= \int_{-\infty}^{\infty}\!d\tau\!\int_{-\infty}^{\tau}\!d\tau'\;
s(\tau)s(\tau')\,
e^{i(\chi a\Omega-\omega)\tau}\,
e^{i(\chi' a\Omega'+\omega)\tau'}.
\end{equation}
For the normalized Gaussian switch, $s(\tau)=\frac{1}{T\sqrt{2\pi}}e^{-\tau^{2}/(2T^{2})}$, this integral has the exact solution (see Appendix~\ref{app:integral}):
\begin{equation}
\mathcal{I}^{\chi\chi'}_{T}(\Omega,\Omega')
= \frac{1}{2}\,
\exp\!\left[-\frac{a^{2}T^{2}}{4}\big(\chi\Omega+\chi'\Omega'\big)^{2}\right]\,
\faddeeva\!\left(T\!\left[\frac{a}{2}\big(\chi\Omega-\chi'\Omega'\big)-\omega\right]\right),
\end{equation}
where $\faddeeva(z)=e^{-z^{2}}\mathrm{erfc}(-iz)$ is the Faddeeva function.

Combining the simplified prefactor with the solution for the integral yields the final, unified expression for the two-photon amplitude:
\begin{empheq}[box=\widefbox]{align}
\mathcal{M}_{\chi\chi'}(\Omega,\Omega')
=&\;\frac{g^{2}a^{2}}{2\hbar^{2}}\,
\chi\chi'\,\Omega\Omega'\;
a^{i(\Omega+\Omega')}\,
\mathcal{N}(\Omega)\mathcal{N}(\Omega')\,
e^{\frac{\chi\pi\Omega+\chi'\pi\Omega'}{2}}
\nonumber\\[2pt]
&\times
\exp\!\left[-\frac{a^{2}T^{2}}{4}\big(\chi\Omega+\chi'\Omega'\big)^{2}\right]\,
\faddeeva\!\left(T\!\left[\frac{a}{2}\big(\chi\Omega-\chi'\Omega'\big)-\omega\right]\right).
\end{empheq}
The specific amplitudes for the RR, LL, RL, and LR channels are obtained by setting $(\chi,\chi')=(+1,+1)$, $(-1,-1)$, $(+1,-1)$, and $(-1,+1)$ respectively.

\section{Evaluation of \texorpdfstring{\(\mathcal{I}(\alpha,\beta)\)}{}} \label{app:integral}
We evaluate
\begin{equation}
\mathcal{I}(\alpha,\beta)=\int_{-\infty}^{\infty}\!d\tau\!\int_{-\infty}^{\tau}\!d\tau'\,s(\tau)s(\tau')\,e^{i\alpha\tau+i\beta\tau'}\,,\qquad s(\tau)=\frac{1}{T\sqrt{2\pi}}\,e^{-\tau^{2}/(2T^{2})}\,.\label{eq:I-app-def}
\end{equation}
Introduce the time-ordered variables \(\Delta\equiv \tau-\tau'\ge0\) and \(s\equiv(\tau+\tau')/2\), so that \(\tau=s+\Delta/2\), \(\tau'=s-\Delta/2\), and \(d\tau\,d\tau'=ds\,d\Delta\). One finds
\begin{equation}
\tau^{2}+\tau'^{2}=(s+\tfrac{\Delta}{2})^{2}+(s-\tfrac{\Delta}{2})^{2}=2s^{2}+\tfrac{\Delta^{2}}{2}\,,\qquad i\alpha\tau+i\beta\tau'=i(\alpha+\beta)s+\tfrac{i}{2}(\alpha-\beta)\Delta\,.
\end{equation}
Hence the integrand factorizes and, after some calculation, it yields
\begin{align}
\mathcal{I}(\alpha,\beta)
&=\frac{1}{2\pi T^{2}}\int_{0}^{\infty}\!d\Delta\,e^{-\Delta^{2}/(4T^{2})}e^{\frac{i}{2}(\alpha-\beta)\Delta}\int_{-\infty}^{\infty}\!ds\,e^{-s^{2}/T^{2}}e^{i(\alpha+\beta)s}\nonumber\\
&=\frac{1}{2\pi T^{2}}\Bigg[\sqrt{\pi}\,T\,e^{-\frac{T^{2}}{4}(\alpha+\beta)^{2}}\Bigg]\int_{0}^{\infty}\!d\Delta\,e^{-\Delta^{2}/(4T^{2})}e^{\frac{i}{2}(\alpha-\beta)\Delta}\,.\label{eq:I-factor}
\end{align}
Set \(\Delta=2Tw\) so \(d\Delta=2T\,dw\) and define \(\kappa\equiv T(\alpha-\beta)\). Then
\begin{equation}
\mathcal{I}(\alpha,\beta)=\frac{e^{-\frac{T^{2}}{4}(\alpha+\beta)^{2}}}{\sqrt{\pi}}\int_{0}^{\infty}\!dw\,e^{-w^{2}}e^{i\kappa w}\,.
\end{equation}
Using the standard half-Gaussian identity (valid for complex \(b\))
\begin{equation}
\int_{0}^{\infty}\!dw\,e^{-w^{2}+2bw}=\frac{\sqrt{\pi}}{2}\,e^{b^{2}}\operatorname{erfc}(-b)\,,
\end{equation}
with \(b=i\kappa/2\), we obtain
\begin{align}
\mathcal{I}(\alpha,\beta)
&=\frac{1}{2}\,e^{-\frac{T^{2}}{4}(\alpha+\beta)^{2}}\,e^{-\kappa^{2}/4}\,\operatorname{erfc}\!\Big(-\tfrac{i\kappa}{2}\Big)\,.\label{eq:I-erfc}
\end{align}
Equivalently, in terms of the Faddeeva function \(\faddeeva(z)=e^{-z^{2}}\operatorname{erfc}(-iz)\), it turns out
\begin{equation}
\boxed{\,\mathcal{I}(\alpha,\beta)=\frac{1}{2}\,\exp\!\Big[-\frac{T^{2}}{4}(\alpha+\beta)^{2}\Big]\;\faddeeva\!\Big(\tfrac{T}{2}(\alpha-\beta)\Big)\,}\label{eq:I-w}
\end{equation}
which is often the most compact form for subsequent frequency-space manipulations. We study some limits in the following:
\begin{enumerate}
    \item Setting \(\alpha=\beta\) gives \(\kappa=0\) and \(\operatorname{erfc}(0)=1\), hence
\(\mathcal{I}(\alpha,\alpha)=\tfrac{1}{2}\,e^{-T^{2}\alpha^{2}}\). Since the Fourier transform of \(\chi\) is \(\tilde{\chi}(\nu)=\exp[-\tfrac{1}{2}T^{2}\nu^{2}]\), this matches the time-ordering identity \(\mathcal{I}(\alpha,\alpha)=\tfrac{1}{2}[\tilde{\chi}(\alpha)]^{2}\).
    \item For large detuning \(|T(\alpha-\beta)|\gg1\), using \(\faddeeva(z)\sim i/(\sqrt{\pi}z)\) yields the algebraic suppression \(\mathcal{I}(\alpha,\beta)\sim \tfrac{i}{\sqrt{\pi}}\,\dfrac{e^{-\frac{T^{2}}{4}(\alpha+\beta)^{2}}}{T(\alpha-\beta)}\).
\end{enumerate}

\section{Distributional limits for the normalized Gaussian switch}
\label{app:limits-Gauss}

We adopt the normalized Gaussian gate
\begin{equation}
s(\tau)=\frac{1}{T\sqrt{2\pi}}\,e^{-\tau^2/(2T^2)},\qquad T>0,
\end{equation}
and define
\begin{equation}
\sigma\equiv\alpha+\beta,\qquad \Delta\equiv\alpha-\beta.
\end{equation}

\subsection*{Exact identity (all $T>0$)}

The time-ordered integral evaluates to
\begin{equation}
\mathcal{I}(\alpha,\beta)
=\int_{-\infty}^{\infty}\!d\tau\!\int_{-\infty}^{\tau}\!d\tau'\,
s(\tau)s(\tau')\,e^{i\alpha\tau+i\beta\tau'}
=\frac{1}{2}\,e^{-\frac{T^2}{4}\sigma^2}\;
\faddeeva\!\Big(\frac{T}{2}\Delta\Big).
\label{eq:I-exact}
\end{equation}

\subsection*{Small-$T$ limit (instantaneous gate)}

The Taylor series of $\faddeeva$ about $z=0$,
\begin{equation}
\faddeeva(z)=1+\frac{2i}{\sqrt{\pi}}\,z - z^2 + \mathcal{O}(z^3),\qquad z=\tfrac{T}{2}\Delta,
\label{eq:w-series}
\end{equation}
together with $e^{-\tfrac{T^2}{4}\sigma^2}=1-\tfrac{T^2}{4}\sigma^2+\mathcal{O}(T^4)$, gives
\begin{align}
\mathcal{I}(\alpha,\beta)
&=\frac{1}{2}\Big(1-\frac{T^2}{4}\sigma^2\Big)\Big(1+\frac{iT}{\sqrt{\pi}}\Delta-\frac{T^2}{4}\Delta^2+\mathcal{O}(T^3)\Big)\nonumber\\
&=\frac{1}{2}+\frac{iT}{2\sqrt{\pi}}\,\Delta-\frac{T^2}{8}\big(\Delta^2+\sigma^2\big)+\mathcal{O}(T^3),
\qquad T\to 0.
\label{eq:I-smallT}
\end{align}
In particular, $\mathcal{I}(\alpha,\beta)\to \tfrac{1}{2}$ as $T\to 0$.

%

\subsection*{Large-$T$ behavior (distributional)}

Introduce the Gaussian $\delta$–sequence
\begin{equation}
\delta_T(\sigma)\equiv \frac{T}{2\sqrt{\pi}}\,e^{-\frac{T^2}{4}\sigma^2}
\;\xrightarrow[T\to\infty]{\ \mathcal S'\ }\;
\delta(\sigma).
\label{eq:deltaT}
\end{equation}
Using the integral representation (for $\operatorname{Im}z>0$, with boundary values by the $+i0^{+}$ prescription)
\begin{equation}
\faddeeva(z)=\frac{i}{\pi}\int_{-\infty}^{\infty}\frac{e^{-t^2}}{\,z-t\,}\,dt,
\label{eq:w-int}
\end{equation}
and the rescaling $t=\tfrac{T}{2}u$, one obtains
\begin{equation}
\frac{1}{2}\,\faddeeva\!\Big(\tfrac{T}{2}\Delta\Big)
=\frac{i}{2\pi}\int_{-\infty}^{\infty}\frac{e^{-\frac{T^2}{4}u^2}}{\Delta-u+i0^{+}}\,du.
\label{eq:w-rescaled}
\end{equation}
As $T\to\infty$, the factor $e^{-\frac{T^2}{4}u^2}$ becomes sharply peaked at $u=0$ with total mass
$\int e^{-\frac{T^2}{4}u^2}\,du=\frac{2\sqrt{\pi}}{T}$. Thus,
\begin{equation}
\frac{1}{2}\,\faddeeva\!\Big(\tfrac{T}{2}\Delta\Big)
= \frac{i}{\sqrt{\pi}\,T}\,\frac{1}{\Delta+i0^{+}}
\;+\;O\!\Big(\frac{1}{T^{3}}\Big).
\label{eq:w-asymp}
\end{equation}
Combining with \eqref{eq:I-exact} and $e^{-\frac{T^{2}}{4}\sigma^{2}}=\frac{2\sqrt{\pi}}{T}\,\delta_T(\sigma)$,
\begin{equation}
\mathcal{I}(\alpha,\beta)
= \frac{2i}{T^{2}}\,\delta_T(\sigma)\,\frac{1}{\Delta+i0^{+}}
\;+\;O\!\Big(\frac{1}{T^{4}}\Big).
\label{eq:I-largeT-asymp}
\end{equation}
In particular, the strict distributional limit is
\begin{equation}
\mathcal{I}(\alpha,\beta)\xrightarrow[T\to\infty]{\ \mathcal S'\ } 0.
\label{eq:I-largeT-zero}
\end{equation}

\subsection*{Mapping to the unified two-photon amplitude}

For the unified amplitude, identify
\begin{equation}
\sigma=a\big(\chi\Omega+\chi'\Omega'\big),\qquad
\Delta=a\big(\chi\Omega-\chi'\Omega'\big)-2\omega.
\end{equation}
In the long-gate regime, the strict limit of $\mathcal{I}^{\chi\chi'}_{T}$ is zero by \eqref{eq:I-largeT-zero}, and the leading term from \eqref{eq:I-largeT-asymp} is
\begin{equation}
\mathcal{I}^{\chi\chi'}_{T}(\Omega,\Omega')
\approx \frac{2i}{T^{2}}\,
\frac{1}{a}\,\delta\!\big(\chi\Omega+\chi'\Omega'\big)\,
\frac{1}{\,a(\chi\Omega-\chi'\Omega')-2\omega+i0^{+}\,}.
\end{equation}

\bibliographystyle{jhep}
\bibliography{UnruhRef}
\end{document}